# Paper-based colorimetric sensor for detection of chloride anions in water using an epoxy-silver nanocomposite

Alfredo Franco[a], Celso Velásquez Ordoñez[b], Miguel Ojeda Martínez[b], María Luisa Ojeda Martínez[b], Enrique Barrera Calva[c], Víctor Rentería Tapia[b*]

[a]Departamento de Física Aplicada, Facultad de Ciencias, Universidad de Cantabria, Santander 39005, Spain
[b]Departamento de Ciencias Naturales y Exactas, Universidad de Guadalajara–Centro Universitario de los Valles, Ameca, Jalisco, 46600, México
[c]Departamento de Ingeniería de Procesos e Hidráulica, Universidad Autónoma Metropolitana–Iztapalapa, A.P. 55-534, Ciudad de México, México

*Corresponding author:
Dr. Víctor Rentería Tapia
Departamento de Ciencias Naturales y Exactas
Centro Universitario de los Valles, Universidad de Guadalajara
Address: Carretera Guadalajara–Ameca Km. 45.5, C.P. 46600, Ameca, Jalisco, México.
Phone: +523757580500
E-mail: victor.renteria@academicos.udg.mx

## Abstract

An epoxy-silver nanocomposite printed on paper was prepared for colorimetric detection of chloride anions ($Cl^-$) in aqueous solution. This paper–based sensor is attractive because can be made following a simple procedure, it shows intense colors, fast naked–eye response and high specificity toward $Cl^-$ detection at concentrations in the 20–400 mM range, far beyond the usual concentrations at which most of the $Cl^-$ colorimetric sensors are limited to. A good linear relationship ($R^2$ =0.9754) between logarithmic $Cl^-$ concentration and the extinction difference at 515 nm was observed with a limit of detection (LOD) of 14 mM. It was also proposed a sensing mechanism to explain the color changes of the sensor (originally yellow–orange color) with the concentration of $Cl^-$: anisotropic silver nanoparticles approaching one another in plasticized resin due to the presence of water (chestnut–brown colored sensor), subsequent interaction between $Cl^-$ and the surface of the silver nanoparticles cause oxidative etching and formation of chain–like aggregates resulting in strong interparticle plasmonic coupling (green–brown colored sensor). This paper–based sensor can distinguish individual $Cl^-$ from other ions such as $F^-$,$OH^-$, $NO_3^-$, $SO_4^{2-}$, $HPO_4^{2-}$,$H_2PO_4^-$,$H^+$, $K^+$, $Na^+$, $NH_4^+$, $Zn^{2+}$ and $Co^{2+}$ as well as in complex mixtures of them. The nanosensor was also tested to recognize $Cl^-$ in seawater and a commercial electrolyte solution, even using volumes as small as 4 μL, suggesting its easy inclusion in portable devices. This novel colorimetric platform is undoubtedly useful for the recognition of $Cl^-$ in environmental and physiological systems.

**Keywords:** Colorimetry; Nanosensor; Silver nanoparticles; Chloride anion; Epoxy resin

**Introduction**

Sensing Cl⁻ in water is a significant topic of current research because they play key roles in physiological and environmental systems [1]. Selective, fast and sensitive detection of such analytes could greatly simplify tests for diseases diagnosis, environmental monitoring and food safety evaluation. In many cases, the chemical sensing requires reaching very low detection limits, in the μM range or less, to evaluate risks for health or the environment [2]. For instance, the permissible limit of Cl⁻ concentration in drinking water is 250 mg/L [1]. Nevertheless, it is also required to recognize and quantify selectively Cl⁻ at high concentrations for medical diagnosis or to assess environmental risks. For example, the detection of chloride anions in sweat at concentrations exceeding 60 mM suggests cystic fibrosis [3]. As well, Cl⁻ at concentrations larger than 250 mg/L may increase the levels of heavy metals in drinking–water [4], and it has also been found very large concentrations of chloride ions (19,000 mg/L) in the composition of oilfield wastewater [5]. The determination of Cl⁻ in aqueous solutions is usually carried out by spectroscopic and electrochemical methods, ion chromatography, surface Raman enhanced scattering and few other techniques reported elsewhere [6]. All these methods are highly sensitive but they are expensive, time–consuming and impractical to perform in–situ medical or environmental tests [7]. So, great efforts have been focused on the visual perception of Cl⁻ in water through a simple, fast, low–cost and portable platform [2].

In this work we explore the performance of a simple, fast, low-cost and portable sensor for colorimetric detection of chloride anions in water based media at high concentrations. There are not many sensing options able to satisfy such criteria, but their development may even boost the use of wearable devices designed to monitor several physiological conditions in people through their own sweat [8].

Few molecular moieties have been designed for the specific colorimetric detection of Cl⁻ based on supramolecular chemistry [9–13]. Despite their good sensitivity and selectivity, they usually involve complicated organic synthesis procedures, water–incompatible properties and their recognition capabilities are restricted to organic liquid phases. This is why the nanostructured plasmonic platforms are being considered as an interesting alternative for the development of sensitive optical sensors enabling naked–eye chloride anions detection [14–17]. In particular, the active agents of the sensor we are reporting here are plasmonic nanoparticles made in just few steps through a bottom-up synthesis procedure.

The plasmonic nanoparticles exhibit large extinction coefficients and remarkable optical properties due to their electrons resonant collective oscillations localized on their surface. The localized surface plasmon resonances (LSPR) greatly depend on the nanoparticles size, shape, composition and surrounding refractive index, any change on these conditions are necessarily translated to a change in the color of any system made of plasmonic nanoparticles, it makes them ideal for label-free sensing and also, because the nanoparticles high surface-volume ratio, ideal for sensing low volume samples [18, 19].

Amongst the plasmonic materials, silver outstands because its sharp and intense LSPR associated spectral extinction bands [20]. Silver is not as chemically stable as gold [20], but this is not an issue when fast sensing responses are the objective, actually the silver chemical reactivity is part of the sensing mechanism behind few sensors [14-16]. Several mechanisms have been proposed to explain their sensing features, such as chemisorption, nanoparticles aggregation, nanoparticles oxidative etching or a combination of all of them. For instance, the recognition and transduction of Cl⁻ in water using silver nanoplates have been associated to chemisorption, involving the formation of silver chloride on these nanoparticles [15]. In this case, the strong surface plasmon spectral band centered at 769 nm is blue–shifted as the concentration of Cl⁻ increases in the range of 0–20 μM. In another example, the interaction between 20 nm diameter starch–capped silver nanoparticles and Cl⁻ in aqueous medium, promotes the formation of silver chloride compounds on the nanoparticles surface, reductions in the size of the nanoparticles and their subsequent aggregation, even if the capping agent is in excess [16]. Specific reactions of halide ions with metal nanostructures result in lower surface charge and a zeta potential decrease, which could initiate the metal nanoparticles aggregation [14]. Under these circumstances, the aggregation leads to both a damping and a broadening of the plasmon resonance spectral band, which extends up to the red–near infrared region due to the near–field interactions of closely spaced metal nanoparticles.

The sensor we are reporting here uses silver nanoparticles as sensing active agent because their optical properties, but also because they can be made from cost-effective bottom-up syntheses procedures. In fact, the synthesis procedure we are reporting in this work uses epoxy resin as all-in-one, solvation, reduction and stabilization reagent, which makes this sensor a unique in its kind, because its nanocomposite synthesis greatly simplifies the sensor fabrication [21, 22]. Moreover, the epoxy/silver nanocomposite is a transparent highly viscous polymeric medium [21] which can be easily printed on cellulose paper to make a cost-effective, flexible and portable sensor.

The sensor reported in this work has been designed for fast sensing of highly concentrated chloride anions in water based samples, the sensor is not designed for sensitive measurements of low chloride concentrations neither for remediation purposes through selective capture of huge ions amounts. Other approaches based on functionalized highly porous platforms are more appropriate for such purposes. In fact, there is a wide variety of highly sensitive and selective colorimetric sensors, based on mesoporous functional materials, for ions capture and environmental remediation [23-28], but such platforms are not as available, cost-effective, flexible and easy to prepare as cellulose paper is.

Recently, cellulose papers impregnated with flexible polymeric materials have been used as cost–effective sensing platforms for analytical and medical chemistry applications [29]. The interest in these platforms comes from their portability, disposability, biodegradability and compatibility with many chemical/biochemical moieties. They allow easy penetration of liquids within their hydrophilic fibers decorated with a variety of functional groups. The 98% α–cellulose filter paper Whatman$^{TM}$ 1 is maybe the most used paper for the elaboration of paper–based colorimetric sensors. It has a smooth surface with no additives, medium flow rate and it allows printing by commercial machines. For instance, a paper–based analytical device was wax–printed to generate hydrophobic barriers and hydrophilic channels to quantify $Cl^-$ in different kinds of water samples, using silver nanoprisms as colorimetric agent [17]. A strong spectral blue shift from 543 to 455 nm and a gradual absorbance decrease was attributed to oxidative $O_2/Cl^-$ etching and transformation of truncated triangles and hexagonal silver nanoparticles into circular disks. Chloride ions preferentially etch the corners and the side faces of silver triangular nanoplates progressing towards their central region, forming disk–like nanoplates, as reported elsewhere [30]. The correlation between the mean color intensity of the nanoparticles and the $Cl^-$ concentration has been used to determine the $Cl^-$ levels in water samples, making use of a smartphone coupled to the paper–based analytical device [17].

Table 1 provides a general overview of the main features of silver plasmonic paper-based colorimetric platforms for sensing chloride anions that we are dealing with in this work. Below we report a low–cost simple synthesis of silver nanoparticles embedded in epoxy resin and their subsequently printing on paper strips, for naked–eye detection of $Cl^-$ in aqueous solution as a function of their concentration. This colorimetric platform was tested for the qualitative detection of chloride ions in complex liquid samples where other anions and cations are present too, even in tap water, seawater and electrolyte solutions. Finally, we propose a mechanism to explain the colorimetric changes associated with the detection of $Cl^-$.

[Insert Table 1]

**Materials and methods**

*Reagents and solutions*

Silver nitrate ($AgNO_3$) and Araldite 506 ™ epoxy resin, both were purchased from Sigma Aldrich. The analytic grade inorganic salts, KF (≥ 99%), KCl, (≥ 99%), KBr (≥ 99%), KI (≥ 99%), $KNO_3$ anhydrous (≥ 99% ), KOH (≥ 97%), $KH_2PO_4$ (99%), $Na_2SO_4$ anhydrous (≥ 99%), NaCl (≥ 99%), $Na_2HPO_4$ (≥ 98), $NH_4OH$ (28 %), NaOH (98%), $Zn(NO_3)_2 \bullet 6H_2O$ (98%), $Co(NO_3)_2 \bullet 6H_2O$ (98%) and HCl (37%) were purchased from Golden Bell and Sigma Aldrich. Solutions at concentrations of $Cl^-$ in the range of 0–400 mM were prepared from their corresponding salts using demineralized water with a resistivity of 18 MΩcm. A $Cl^-$ solution at 60 mM mixed with anions of $NO_3^-$, $SO_4^-$ and $OH^-$ (named $A^-$), was prepared from the NaCl, $KNO_3$, $Na_2SO_4$, and $NH_4OH$ compounds, respectively. The individual concentration of $A^-$ was fixed at 40 mM (molar concentration ratio $r$ = [Cl$^-$]/[A$^-$] of 1.5). All the reagents were used as received without any purification procedure.

*Synthesis of the silver nanoparticles*

Silver nanoparticles were synthesized in epoxy resin as follows. The silver salt precursor (AgNO$_3$) was mixed in the epoxy resin (Araldite 506 ™) in a glass reactor. The mixture was stirred slowly at 60°C for 24 hours. In this mixture, the estimated concentration of silver ions was 6.8 mM. The epoxy resin works both as dispersing and reducing agent of silver ions [22]. The mixture looks yellow after few minutes, but it becomes dark red as the reaction proceeds, because of the formation of silver nanoparticles. Fig. 1 shows a scheme that illustrates the synthesis and structure of the nanocomposite. First, the resin polar groups (mainly hydroxy groups) solvate the AgNO$_3$ salt by means of oxygen interactions (R-O-Ag$^+$), then the resin is thermo-oxidated to produce CH$_3$·radicals [31] and subsequently silver atoms (Ag$^0$), as described by equation (1) in Fig. 1. Silver clusters (Ag$^0_m$) are formed after the aggregation of m reduced silver atoms, as described by equation (2) in Fig. 1. Finally, after silver atoms addition, anisotropic silver nanoparticles (Ag$^0_n$) are formed, as described by equation (3) in Fig. 1.

[Insert Fig. 1]

*Paper-based epoxy-silver nanosensor fabrication*

As soon as the silver/resin nanocomposite reaches room temperature, it was roll–printed on Whatman™ 1 filter paper. The size of the filter paper samples was 2.2 x 8.1 cm², and the printed area was close to 0.7 x 8.1 cm². After printing, the filter paper was dried at 90°C for 1 h in air atmosphere. The color of the printed area looks yellow–orange after the thermal treatment. The nanocomposite red color turns yellow-orange due to the paper-nanoparticles optical scattering. At this point, the paper is ready to be cut in the form of strips for its use as a sensor. The typical strips size was 0.5 x 2.0 cm². The sensing strips can be stored for later use by simply keeping them protected from light and humidity at room temperature.

*Determination of chlorides by the paper-based epoxy-silver nanosensor*

The sensor preparation conditions as well as the time the sensor is exposed to the anionic solutions were optimized for the colorimetric measurements. The paper–based sensor was immersed in aqueous single–anion solutions for 10–12 seconds at room temperature ($\sim$35 $^0$C). Intense colors were visualized by the naked–eye in this period; such colors depend on the anions nature and its concentration. After the immersion and just before the optical measurement, the sensor was placed on absorbent paper for 5 minutes. However, in the case of chlorides mixed with other anions, it was necessary to wait for at least 15 minutes before carring out the measurement, because the mixed ions delay the appearance of the characteristic color associated to Cl$^-$.

*Instrumentation*

The optical extinction was measured by diffuse reflectance spectroscopy (DRS) using an integrating sphere coupled to an UV–Vis–NIR Shimadzu UV–3600 spectrophotometer. The samples were also analyzed by scanning electron microscopy (SEM) with energy–dispersive X–ray spectroscopy (EDS) in a JEOL JSM 7800F microscope operated at 15 kV, to investigate the morphology and to identify silver nanoparticles on the paper-based sensor.

**Results and discussion**

*Optical and morphological characterization*

Fig. 2a shows the UV–Vis extinction spectrum of the Whatman™ paper with and without the epoxy/silver nanocomposite. A picture of this printed paper is also shown in the upper right inset of the figure. An intense broad extinction band with its maximum located at 425 nm was observed for the paper printed with the nanocomposite, whereas no peak was observed for the non–printed paper. Fig. 2b shows the extinction of the printed paper as a function of the aging time. It is observed a gradual blue–shift from 435 to 425 nm in a few days. Although the extinction strongly decreases for longer periods, the paper sensor was still sensitive to the analytes. In the inset of this figure, we can also observe that the extinction at 425 nm from the printed paper is practically constant for short aging times.

[Insert Fig. 2]

SEM images of the printed paper were obtained to know the morphology of the silver nanoparticles on the substrate. Because the optical properties of the silver nanoparticles are highly dependent on their shapes, sizes and interparticle separation. The SEM images reveal island–like resin regions containing silver nanoparticles, in several sizes and shapes, dispersed along the islands, but also with particles very close each other (Fig. 3a). The geometry of the silver nanoparticles corresponds mainly to both truncated polyhedrons with different planar facets, sharp edges and corners as well as some rounded particles, most of them separated and a few in close contact (Fig. 3b). The single nanoparticles size ranges from 50 up to 250 nm with interparticle separation distances of several hundred nanometers (~500 nm).

[Insert Fig. 3]

The dependence on the shape of the nanoparticles is stronger in the blue, green and red spectral regions, for spherical and roughly spherical shaped nanoparticles, decahedral and triangular truncated pyramids, respectively [32]. On the other hand, the scattering predominates when the size of the particles increases (> 40 nm), the resonance shifts its position to longer wavelengths, but damping and broadening effects are observed too [33–34]. Similarly, as the particles reach a multipolar regime of excitations, the extinction spectra reveal it by means of broadening and the appearance of several peaks. Thus, it is expected that a mixture of spherical and faceted silver particles, such as the one we report, exhibit a broad extinction band [35]. Therefore, the wide extinction band extending along the UV–Vis region in Fig. 2a is mainly attributed to mixtures of single silver truncated nanopolyhedrons, small rounded particles and a few particles in close contact.

In order to confirm the presence of silver in the printed paper substrate, EDS measurements were used. Fig. 4 shows the EDS spectrum of the printed paper, where silver is identified besides other non–silver traces coming from the own paper composition.

[Insert Fig. 4]

*Colorimetric tests*

Whatman™ 1 paper printed with the epoxy–silver nanocomposite was used as a paper–based colorimetric sensor for a fast determination of Cl⁻ in aqueous medium, according to the following simple procedure. The printed paper strip was immersed in the aqueous solution containing Cl⁻ for 10 to 12 s. Intense green–brown color hues appear during this period at different concentrations of Cl⁻ in the range of 20–400 mM, easily observed by the naked eye. After the immersion, the sensor is placed on absorbent paper for at least 3–5 min, after while the sensor is ready for the optical reading (Fig. 5).

Fig. 6 shows the UV–Vis extinction spectra of the sensor after its immersion in aqueous medium with a few Cl⁻ concentrations ranging from 0 to 400 mM. At a concentration of 0 mM of Cl⁻ a spectral shoulder at 515 nm becomes evident as well as the broadening of the surface plasmon resonance band. However, after the sensor immersion in aqueous solutions containing Cl⁻ at concentrations from 7 to 400 mM, the surface plasmon resonance band at 425 nm is blue–shifted to 415 nm. The intensity of the shoulder at 515 nm gradually diminishes in comparison to the one registered for the identification of water. For Cl⁻ concentrations in the 180–400 mM range no more significant changes in the optical extinction were observed.

The paper–based sensor exhibits different color hues and saturation levels, distinguishable by the naked eye, as a function of the Cl⁻ concentration, in the range of 0–400 mM, as the inset in Fig. 6 shows. The original yellow–orange color of the paper–based sensor becomes chestnut–brown for Cl⁻ concentrations in the range of 0–7 mM. The color turns green–brown for 20–120 mM concentrations and the color intensity decreases for the 180–400 mM range. After these tests, the lowest detection limit distinguishable by the naked eye is estimated to be around 20 mM (color green-brown).

The extinction difference ($\Delta I$) at 515 nm wavelength is a linear function ($R^2 = 0.9754$, $n = 3$) of the logarithmic concentration of Cl⁻ in the range of 20 to 400 mM. It shows a LOD equal to 14 mM (Fig. 7). The extinction difference was determined as the difference between the extinction of silver nanoparticles in absence ($I_0$) and presence ($I$) of Cl⁻, respectively. While the LOD was calculated from the concentration $x$ at which $log(x) = 3\sigma/s$, where $\sigma$ is the standard diviation of the blank (< 3%) and $s$ the slope of the calibration graph in the linear range, respectively [23]. The correlation suggests that the sensor can be used for the quantification of single Cl⁻ in aqueous medium.

SEM images of the sensor were obtained after its immersion in aqueous solutions with Cl⁻ at 0 and 120 mM concentrations (Fig. 8). For 0 mM the images at 10,000x show faceted and elongated silver nanoparticles agglomerates (Fig. 8a). Elongated silver nanoparticles in approaching process were also observed (Fig. 8b). The effective diameter of these close together particles and their interparticle distance was estimated in the range of 100–500 nm with a separating distance around 150 nm, respectively. A similar morphology has been reported elsewhere where the plasmonic coupling, due to quasi–spherical silver nanoparticles and faceted particles dispersed in aqueous medium, originates a spectral shoulder around 540 nm [36]. Consequently, the color change from yellow–orange to chestnut–brown, related to both the spectral shoulder at 515 nm and the broadening (Fig. 6), is due to the increment in the size of the particles and their subsequent agglomeration, resulting in plasmonic coupling induced by water.

In presence of Cl⁻, the silver nanoparticles tend to form chain–like agglomerates (Fig. 8c). These agglomerates consist of many very close together individual rounded small particles, from 50 to 100 nm in diameter (Fig. 8d). It is interesting because as the chain interparticle distance shortens the near–field coupling becomes more relevant [37]. Thus, the silver nanoparticles–chloride ions interaction induces both a spectral blue shift from 425 to 415 nm and an optical extinction at longer wavelengths, properties related with the color change from yellow–orange to green–brown. It is noteworthy that the island–like resin regions containing micro/nanometric silver nanoparticles (Fig. 3a) were fragmented in linear agglomerates, few hundreds of micrometers in length, and spread out along the paper (Fig. 8d).

On the other hand, the gradual decrease in the intensity of the spectral shoulder located at 515 nm in presence of Cl⁻ could be explained in terms of a morphological transformation carried out by oxidative etching of the silver nanoparticles, as noted below.

[Insert Fig. 8]

In general, under natural conditions, the sensor lasts around 40 minutes to reach its maximum optical extinction at 415 nm. The sensor keeps such extinction stable for around 20 minutes more, after which its optical extinction slowly decreases. It lasts around 90 hours to come back to its initial extinction value. However, for immediate response sensors, as it is the case, this dynamic response is not an issue.

*Cl⁻ detection selectivity*

In order to test the selectivity of the paper–based sensor toward Cl⁻, we studied its colorimetric response toward several solutions in absence and presence of individual ions such as $F^-$, $NO_3^-$, $OH^-$, $H_2PO_4^-$, $SO_4^{2-}$, $HPO_4^{2-}$, $NH_4^+$, $Zn^{2+}$ and $Co^{2+}$ at the fixed concentration of 120 mM in all cases. According to the results, no relevant color changes were observed for Cl⁻ in absence (Fig. 9a) or in presence of interfering ions (Fig. 9b). The colorimetric tests of the sensor toward Cl⁻ in presence of a variety of ions suggest good selectivity, since only Cl⁻ induce a distinct color change from yellow-orange to green-brown, resulting a blue-shift from ~425 to 415 nm and a decrease of the extinction at 515 nm. Besides, the monovalent metal cations from chloride salts (KCl, NaCl) and HCl in aqueous solution at 120 mM give rise to the same color on the paper–based sensor and good selectivity after its immersion in these aqueous solutions, as it is shown in the Fig. 9c. The Cl⁻ concentration and the concentration of the interfering species used for the selectivity study is much higher than the one found in typical aqueous samples from environment or physiological systems [17].

[Insert Fig. 9]

These results suggest that the sensor may distinguish chloride from other individual ions, or a mixture of them, in a broad range of concentrations. Figure 10a shows some representative examples of two mixtures with several anions. In particular, it was considered the case of individual Cl⁻ at 60 mM, the mixture of Cl⁻ with anions of $NO_3^-$, $SO_4^-$ and $OH^-$ ($Cl^- + A^-$) and the same mixture with addition of $F^-$ ($Cl^- + A^- + F^-$). It was found that when 10 μmol of A⁻ are mixed to the Cl⁻ solution, the surface plasmon resonance has a red shift from 415 to 427 nm and a slight damping in the extinction. The addition of 10 μmol of F⁻ to this same mixture causes more damping and a distinctive shoulder at 545 nm in the optical extinction. However, the feature green–brown color due to Cl⁻ was conserved in both mixtures as the inset of the Fig. 10a shows.

The colorimetric paper–based sensor can be used to detect Cl⁻ in different types of water contained mixed ions. Fig. 10b shows an example of several tests carried out to identify chlorides in sea water (([Cl⁻] = 34000 mg/L) and in a commercial electrolyte solution ([Cl⁻] = 17410 mg/L). The tests were also carried out in bi–distilled water ([Cl⁻] < 0.5 mg/L) and tap water ([Cl⁻] < 50 mg/L). The identification of Cl⁻ was positive for seawater and in the commercial electrolyte solution. The test was successful even testing small sample volumes (4 μL) deposited on the sensor. The recognition of Cl⁻ in bi–distilled and tap water was negative because of the high detection limit of the sensor ( 14 mM). Therefore, this colorimetric platform is interesting to be used for the naked-eye recognition of Cl⁻ at high concentration ([Cl⁻] ≥ 20 mM) in several kinds of aqueous solutions, some of them relevant in environmental and human health inspections.

*Detection mechanism*

In this section, we propose a sensing mechanism behind the sensor recognition of $H_2O$ and $Cl^-$, considering the information obtained from the UV–Vis and SEM analyses, and also taking into account previously reported data.

Water can be quickly adsorbed on the paper and interacts with the resin structure via hydrogen–bonding, leading to a plasticizing effect and a reduction in the viscosity of the resin [38–40]. Under these conditions, high mobility of the epoxy resin without cure is expected on the cellulose fibers due to the presence of water. Additionally, a chemical bonding Ag-O like between silver nanoparticles and epoxy networks via hydroxyl groups of the resin has been already reported [41]. Nevertheless, silver nanoparticles embedded in the plasticized resin could be released and migrate through the cellulose fibers, due to a possible weakening of the Ag–O bonding and to the plasticizing effect of the resin. Thus, water could lead to neutralize the silver nanoparticles surface charge, to decrease the zeta potential, and in consequence to induce the aggregation of metal particles. In the end, these physical–chemical changes translate into a plasmonic coupling. The plasmonic coupling between these silver nanoparticles causes a strong color change from yellow–orange to chestnut–brown in presence of water; it gives also rise to a spectral shoulder at 515 nm and to an extinction band broadening, as it is shown in Fig. 6.

On the other hand, the interaction between silver nanoparticles and chloride ions leads to the formation of chloride compounds adsorbed onto the surface of these particles, which have low solubility in water [14]. In particular, the AgCl solubility product constant ($k_{sp}$) is $1.76 \times 10^{-10}$, it suggests that the adsorption of chloride on the silver nanoparticles surface is thermodynamically favored and it occurs rapidly due to the Ag–chloride high-affinity bond [14,15]. Moreover, the chemisorption of halides on metal nanoparticles is possible even in presence of capping agents like citrate [42], CTAB bilayers [43], or starch acting as both steric and protector agents of silver nanoparticles [16]. It means that the adsorption of chlorides on the surface of silver nanoparticles is possible even if they are embedded in epoxy resin and printed on paper.

In contrast, silver compounds with high $k_{sp}$ values would result in a weak Ag–anion interaction and therefore, the colorimetric changes toward its anions are similar to the case of water due to larger solubility. For example, the $k_{sp}$ values of AgF and AgNO$_3$ are 182 g/100 ml water and 122 g/ 100 ml water, respectively [15]. The color of the sensor after immersion in F$^-$ and NO$_3^-$ solutions is very different from the one in Cl$^-$ but similar to the water case. This is the reason why a distinguishable color is observed in the paper–based sensor for chlorides in comparison with others anions (see Fig. 9). The selectivity toward Cl$^-$ is determined by the solubility of their silver compounds adsorbed on the surface of the silver nanoparticles. Hence, distinguishable recognition between halides will proceed only for silver compounds with very low water solubility depending on their small $k_{sp}$ values. In Fig. 11, the UV-Vis extinction spectra and colorimetric changes corresponding to the detection of F$^-$, Cl$^-$, Br$^-$ and I$^-$ at 120 mM, confirm this last asseveration.

Additionally, the metal nanoparticles agglomerated in presence of water exposed to nucleophilic moieties such as Cl$^-$ may induce morphological transformations via oxidative etching, that would lead to a specific plasmonic coupling, as suggested by the results on halide–triggered shape transformation and aggregation of silver nanoparticles reported elsewhere [17, 30]. The next reaction has been reported to describe the oxidation of silver nanoparticles by chloride ions, giving rise to a shape transformation from silver truncated nanotriangles and nanohexagons to silver circular nanodisks in a paper–based colorimetric sensor [17]:

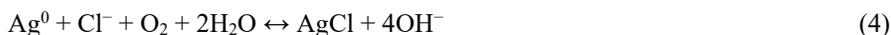

$$Ag^0 + Cl^- + O_2 + 2H_2O \leftrightarrow AgCl + 4OH^- \qquad (4)$$

The chloride–induced silver nanoparticles shape transformations are selectively initiated on the corners of triangular nanoplates, progressing toward the central region and resulting in the formation of disk–like shapes [30]. Similarly, it has been reported that the oxidation of the polyhedral silver nanoparticles deposited on paper is favored where the silver atoms can be easily etched, due to the high surface energy at their sharp edges and corners [17]. In our case, a similar oxidative etching process of the silver nanoparticles upon addition of Cl$^-$ could explain the shape transformation from polyhedral nanoparticles to rounded shapes and shorter particle sizes, as Fig. 8 shows. Besides, it has been reported that the formation of halide compounds on the surface of the metal nanoparticles would lead to the destabilization of the electronic charge on the metal surface and to increase the Van der Waals attractive forces among these nanoparticles [42].

Consequently, the aggregation of oxidized metal particles very close together plays a dominant role in the colorimetric recognition of chloride ions by the epoxy-silver nanocomposite. As noted above, the silver nanoparticles can move through the paper because the epoxy resin is not cured and due to its plasticization caused by water. This behavior is similar to that one reported for gold nanoparticles in plasticized PVC membranes, where the nanoparticles aggregation is induced by the iodide adsorption on their surfaces and their high mobility through the plasticized PVC [44].

So that, the gradual decrease of the spectral shoulder located at 515 nm after the addition of $Cl^-$ is due to the shape transformation from silver nanopolyhedrons to chain–like aggregates constituted of small rounded and oxidized particles. The chain–like aggregates exhibit an optical extinction with a maximum at 415 nm, which it extends at longer wavelengths as shown in Fig. 6.

Regarding the chain–like aggregation, it is interesting that the structure of the aggregates depends on the chloride compounds formed on the surface of the silver nanoparticles [16]. They could be the consequence of the different molecular geometries of the AgCl and $AgCl_2^-$ ionic pairs. The $AgCl_2^-$ linear structure could orient the aggregates in the form of chain–like structures, while the AgCl non–linear structure would yield bulky–like structure [16]. Therefore, probably in our system the $AgCl_2^-$ moieties are adsorbed on the surface of the silver nanoparticles, and they induce the formation of the unique structures observed in Fig. 8c. Nevertheless, it should also be considered that the nanoparticles could be spread out long distances on the cross-linked cellulose fibers by means of the linear agglomerates of plasticized resin, which lead to the formation of the chain–like structures (Fig. 8d).

On the other hand, the chemisorption of $H^+$, $Na^+$, $K^+$, $NH_4^+$, $Zn^{2+}$ and $Co^{2+}$, on the silver nanoparticles surface is not favored, and therefore it could not modify the morphology of the nanoparticles. This effect has been explained by the high standard reduction potential of the silver nanoparticles in which none of these moieties could lead to a significant oxidation [17].

Fig. 12 summarizes all the main mechanisms involved in the interactions between $Cl^-$ anions and the sensor epoxy/silver nanocomposite. Fig. 12a shows a system of anisotropic silver nanoparticles homogenously distributed along the epoxy resin printed on the paper. Fig. 12b illustrates how the silver nanoparticles migrate to form aggregates after water addition, due to the resin plasticizing effect. Fig. 12c shows that, in presence of oxygen and chloride anions, the silver nanoparticles change their size and shape because etching. The silver nanoparticles also get covered by a silver chloride shell made according to equation (4), which describes the interactions of single silver atoms with water, oxygen and chloride anions. Equation (5) in Fig. 12 generalizes equation (4) to describe the overall interactions occurring in a system of silver nanoparticles printed on the sensor paper.

[Insert Fig. 12]

*Key sensor features*

Table 2 compares the sensor performance with other colorimetric sensors reported for $Cl^-$ detection using different nanocomposites. Table 2 follows a similar format of a table published elsewhere to compare the features of several halide anions sensors [2]. The sensor reported in this work shares few advantages with other similar sensors: its fabrication is not complex, its cellulose paper platform is highly versatile and it provides clear colorimetric readouts. Although this sensor is not an option for highly sensitive and selective chloride anions sensing, it is highly remarkable that this epoxy/silver nanocomposite based sensor fills two gaps in the state-of-the-art: the nanocomposite synthesis is really easier in comparison to other sensors and it works well at high chloride anions concentrations where other sensors are out of their linear detection range. Besides, this sensor does not need to pre-process the samples, it also makes the sensor suitable for sensing low volume samples.

[Insert Table 2]

## Conclusions

A paper–based sensor was prepared, using an epoxy/silver nanocomposite for the naked–eye recognition of Cl⁻ at various concentrations in aqueous solution. In contrast to other sensors, this sensor can be fabricated in just few simple steps because the epoxy resin ability to perform by itself as solvating, reducing and stabilizing reagent. It is also remarkable the sensor capability to work properly without any kind of sample pre-processing, it makes this sensor ideal for cost-effective in-situ measurements. Besides, the sensor linear response range extends up to 29,820 mg/L, far beyond most of the similar sensors capabilities. Intense colors, fast and selective response to Cl⁻ were achieved even in presence of mixtures of different anions and cations. The sensor lowest limit of detection was found to be 1,043 mg/L. Morphological transformations from truncated polyhedrons to chain–like aggregates of the silver nanoparticles result in strong plasmonic coupling, which is associated with the sensor colorimetric changes triggered by the presence of Cl⁻. The molecular interaction of water with the resin structure, the transport properties of the resin on the paper, as well as the low solubility of the silver compounds adsorbed on the silver nanoparticles, were proposed as key factors in the selective recognition of Cl⁻. This explains why the sensor can discriminate Cl⁻ anions from a mixture of other anions in aqueous solutions and even in environmental systems. We expect that the results reported in this work boost the development of physiological and environmental portable devices for the detection of high concentrations of chloride anions in aqueous solutions.

## Authors' contributions

Alfredo Franco: Verification, Methodology, Writing-Review and Editing; Celso Velázquez: Verification; Miguel Ojeda: Formal Analysis; María Luisa Ojeda: Verification; Enrique Barrera: Writing-Review and Editing; Víctor Rentería: Conceptualization, Project administration, Investigation.

## Funding

Víctor Rentería gratefully thanks to the PROSNI program from Universidad de Guadalajara for the partial grant of this investigation.

## Acknowledgments

The authors are thankful to Samuel Tehucanero (IFUNAM) by SEM measurements.

## Conflict of interest

The authors declare no competing interests.

**Table captions**

**Table 1** Aim, approach, transduction process, active agent, reagents and platform of the studied sensor in comparison to other reported alternatives.

**Table 2** Key features of the sensor in comparison to the performance of other reported metal nanoparticle based sensors for colorimetric detection of chloride anions in liquid samples.

**Table 1**

| | Features | Alternatives |
|---|---|---|
| **Aim** | | |
| To detect chloride anions in liquid samples at high concentrations. [1, 3-5] | High concentrations in sweat are related to cystic fibrosis. High concentrations in water are related to water pollution. | To capture large quantities of chloride anions for their removal from sweat or water is an alternative aim out of the scope of this work. |
| **Approach** | | |
| Colorimetric sensing [2, 7] | Fast response. Cost-effective. Easy interpretation. Free of readout instrumentation. | Alternative approaches may be more sensitive, but they are more expensive, time-consuming, need extra instrumentation and they are less portable friendly. |
| **Transduction process** | | |
| Localized Surface Plasmon Resonance (LSPR) [14-20] | Metal nanoparticles as active agents. Active agents with large surface-volume ratios. High sensitivity to the shape and size of the active agents as well as to their surrounding refractive index. Ideal for samples in low amount. Metal nanoparticles are more photostable than organic molecules and their extinction coefficients are higher. Label-free sensing. | Other processes may be more selective, but they need to use organic chromophores and fluorophores, which generally need complicated synthesis procedures, they are less hydrophilic and less photostable than metal nanoparticles. |
| **Active agent** | | |
| Silver nanoparticles [14-17, 20] | The silver LSPR bands are sharper and more intense than gold LSPR bands. Easy bottom-up synthesis. Cost-effective. | Other metal nanoparticles may be more chemically stable, but they are more expensive than silver and their LSPR bands are broader and less intense. |

| | Strong affinity between chloride anions and silver. Detection of chloride anions by the modification of the nanoparticles surrounding refractive index, due to nanoparticles chemisorption, aggregation and oxidative etching. | |
|---|---|---|
| **Reagents** | | |
| Epoxy resin [21, 22] | One step chemical synthesis of silver nanoparticles. All-in-one solvation, reduction and stabilization reagent. Transparent and highly viscous polymeric medium. Printing friendly. | There are other alternatives, but generally related to more complicated and expensive synthesis procedures. |
| **Platform** | | |
| Paper [17, 23-30] | Flexible solid. Portable. Cost-effective. Liquid samples easy penetration. Printing friendly. | Other alternatives include high surface-volume ratio platforms such as mesoporous and metal-organic frameworks, these are ideal for tailored functionalization and great for the capture of large variety of ions, but they are not as flexible and available as paper is. |

**Table 2**

| Nanoparticles | Nanoparticles synthesis | Platform | Samples | Detection principle | Linear determination range | Limit of detection | Reference |
|---|---|---|---|---|---|---|---|
| Hybrid organic-inorganic gold nanoparticles | *Solvating agent*: Water<br>*Reducing agent*: $C_6H_8O_6$<br>*Stabilizer agent*: Organic ligand<br>*Additional reagents*: Yes | Liquid dispersion | River, city and sea water<br>*Pre-processing*: No | UV-visible absorption depends on the $Cl^-$ anions captured by the nanoparticles organic moieties | Up to 1.8 mg/L | 0.10 µg/L | [45] |
| Silver triangle nanoplates | *Solvating agent*: Water<br>*Reducing agent*: $NaBH_4$<br>*Stabilizer agent*: Poly-vinylpyrrolidone and $C_6H_5O_7Na_3$<br>*Additional reagents*: Yes | Paper | Water, pharmaceuticals and tomato juice<br>*Pre-processing*: Yes | Color depends on the spatial separation of nanoparticles due to their interaction with $Cl_2$ gas produced by $Cl^-$ anions | 0.1 – 1.5 mg/L | 0.04 mg/L | [46] |
| Silver triangle nanoplates | *Solvating agent*: Water<br>*Reducing agent*: $NaBH_4$<br>*Stabilizer agent*: Poly-vinylpyrrolidone and $C_6H_5O_7Na_3$<br>*Additional reagents*: Yes | Paper | Saline solutions<br>*Pre-processing*: Yes | Color depends on the spatial separation of nanoparticles due to their interaction with $Cl_2$ gas produced by $Cl^-$ anions | 0.1 – 2 mg/L | 0.03 mg/L | [47] |
| Silver triangle nanoplates | *Solvating agent*: Water<br>*Reducing agent*: $NaBH_4$ | Paper | Water, pharmaceuticals and food | Color depends on the spatial separation of nanoparticles due to their | 0.03 – 0.3 mg/L | 0.01 mg/L | [48] |
| | *Stabilizer agent*: Poly-vinylpyrrolidone and $C_6H_5O_7Na_3$ *Additional reagents*: Yes | | *Pre-processing*: Yes | interaction with $Cl_2$ gas produced by $Cl^-$ anions | | | |
| Silver triangle nanoplates | *Solvating agent*: Water *Reducing agent*: $NaBH_4$ *Stabilizer agent*: Poly-vinylpyrrolidone and $C_6H_5O_7Na_3$ *Additional reagents*: Yes | Paper | Water and food *Pre-processing*: Yes | Color depends on the spatial separation of nanoparticles due to their interaction with $Cl_2$ gas produced by $Cl^-$ anions | N.A. | 0.04 mg/L | [49] |
| Silver nanoprisms | *Solvating agent*: Water *Reducing agent*: $NaBH_4$ *Stabilizer agent*: Starch *Additional reagents*: Yes | Paper | Drinking, tap and ground water *Pre-processing*: No | Color depends on the oxidative etching of nanoparticles induced by $O_2/Cl^-$ | 10 – 1000 mg/L | 1.3 mg/L | [17] |
| Silver nanoparticles | *Solvating agent*: Water *Reducing agent*: $NaBH_4$ *Stabilizer agent*: Starch *Additional reagents*: Yes | Paper | Mineral water *Pre-processing*: No | Color depends on AgCl formation after nanoparticles oxidation with $H_2O_2$ | 25 – 1000 mg/L | 2 mg/L | [50] |
| Silver nanoparticles | *Solvating agent*: Anhydrous ethanol and ultrapure water *Reducing agent*: Chloride ions *Stabilizer agent*: Poly-vinylpyrrolidone | Liquid dispersion | Fuel ethanol *Pre-processing*: No | UV-visible absorption depends on the photochemical generation of nanoparticles in presence of Ag(I) and poly-vinylpyrrolidone | 0.05 – 0.8 mg/L | 0.012 mg/L | [51] |
| | | | | *Additional reagents*: No | | | |
| Silver nanoclusters | *Solvating agent*: 4-(2-hydroxyethyl)-1-piperazineethane-sulfonic acid *Reducing agent*: Formaldehyde *Stabilizer agent*: Poly-ethyleneimine *Additional reagents*: No | Liquid dispersion | Water *Pre-processing*: No | Color depends on the oxidative etching of nanoparticles and their aggregation | 0.02 – 0.89 / 1.8 – 14.2 mg/L | 0.007 – 0.7 mg/L | [52] |
| Silver nanoprisms | *Solvating agent*: Water *Reducing agent*: $NaBH_4$ and $C_6H_5O_7Na_3$ *Stabilizer agent*: Aminopropyl-triethoxysilane *Additional reagents*: Yes | Quartz slides | Water *Pre-processing*: No | Color depends on the etching of nanoprisms to become nanodisks | 10.63 – 212.6 mg/L | 10.63 mg/L | [53] |
| Silver nanoprisms | *Solvating, reducing and stabilizer agent*: Epoxy resin (Araldite 506 $^{TM}$) *Additional reagents*: No | Paper | Water with high $Cl^-$ concentrations, tap water, seawater and electrolyte solutions. *Pre-processing*: No | Color depends on the aggregation, oxidation and etching of the nanoprisms | 1,491 – 29,820 mg/L | 1,043 mg/L | This study |

**Figure captions**

**Fig. 1** Scheme illustrating the synthesis and structure of the silver nanocomposite. The resin polar groups are indicated in red, while the resin groups responsible of the $CH_3^{\bullet}$ radicals are indicated in purple. Equation (1) describes the reduction of silver ions by the $CH_3^{\bullet}$ radicals. Equation (2) describes the silver clusters formation by m reduced silver atoms. Equation (3) describes the anisotropic silver nanoparticles formation

**Fig. 2 a** UV–Vis extinction spectra of the paper without and with the epoxy/silver nanocomposite. The inset is a picture of the printed paper. **b** Extinction spectra of the printed paper as a function of aging time. The extinction of the printed paper is practically constant for short aging times (inset)

**Fig. 3** SEM images of the epoxy/silver nanocomposite printed on the surface of Whatman™ 1 paper. **a** Island–like resin regions containing silver nanoparticles. **b** Mixtures of silver truncated polyhedron nanoparticles with facets, edges and corners, rounded particles and some particles in close contact

**Fig. 4** EDS spectrum of the paper–based sensor, where silver is identified. The inset is a SEM image where the spectrum was obtained from

**Fig. 5** Colorimetric test procedure for the identification of $Cl^-$ in aqueous medium. **a** Paper–based sensor before immersion. **b** Immersion of the sensor in an aqueous solution containing $Cl^-$ at 120 mM. **c** Paper–based sensor after the immersion

**Fig. 6** UV–Vis extinction spectra and pictures of the paper–based sensor (inset) before and after its immersion in $Cl^-$ solution at different concentrations

**Fig. 7** Plot of $\Delta I$ at 515 nm versus $Cl^-$ concentration. Inset: $\Delta I$ versus logarithmic $Cl^-$ concentration calibration graph, in the range of 20–400 mM

**Fig. 8** SEM image of silver nanoparticles on the paper–based sensor in presence of water at 10,000x magnification. **a** Agglomerated silver nanoparticles and **b** elongated silver nanoparticles in approaching process. Chain–like aggregates of silver nanoparticles on the paper–based sensor in presence of $Cl^-$ at **c** 9,000x; **d** 55,000x magnification

**Fig. 9** Pictures of paper–based sensor colorimetric response toward: **a** $Cl^-$ and other ions without mixing. **b** $Cl^-$ with interfering ions. **c** $Cl^-$ coming from potassium and sodium salts and HCl. The concentration of $Cl^-$ was 120 mM in all cases. The colorimetric response in presence of water is registered as reference.

**Fig. 10** UV–Vis extinction spectra and photographic pictures of the colorimetric identification of chloride ions. **a** Single $Cl^-$ in demineralized water at 60 mM, with mixed ions $(Cl^- + A^-)$ and with the addition of $A^-$ and $F^-$ $(Cl^- + A^- + F^-)$. **b** Single $Cl^-$ in demineralized water at 120 mM, $Cl^-$ in seawater and the electrolyte solution. No recognition toward chloride ions was possible in bi–distilled and tap water

**Fig. 11** UV-Vis extinction spectra corresponding to the selective detection of $F^-$, $Cl^-$, $Br^-$ and $I^-$ depending on the $k_{sp}$ of their silver compounds: AgF (182 g/100 ml), AgCl ($1.76 \times 10^{-10}$), AgBr ($5.32 \times 10^{-13}$), AgI ($8.49 \times 10^{-17}$). Pictures of paper–based sensor colorimetric response toward halides are also shown (inset)

**Fig. 12** Graphical representation of the key interaction mechanisms in the sensor between the chloride anions and the epoxy/silver nanocomposite. **a** Nanocomposite printed on the paper. **b** The silver nanoparticles migrate to form aggregates after water addition. **c** Oxidative etching of the anisotropic silver nanoparticles into smaller particles by $Cl^-$ and silver chloride shell on the nanoparticles. Equation (4) describes the single silver atom interaction with water, oxygen and chloride anions, while equation (5) describes the same kind of interaction but for a system of silver nanoparticles printed on a paper.

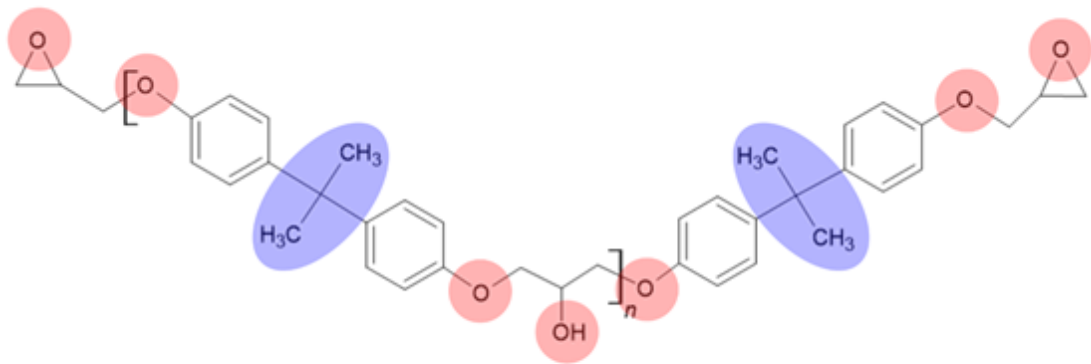

(1) 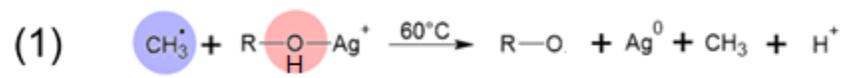

(2) 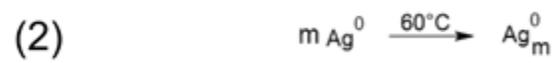

(3) 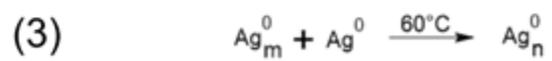

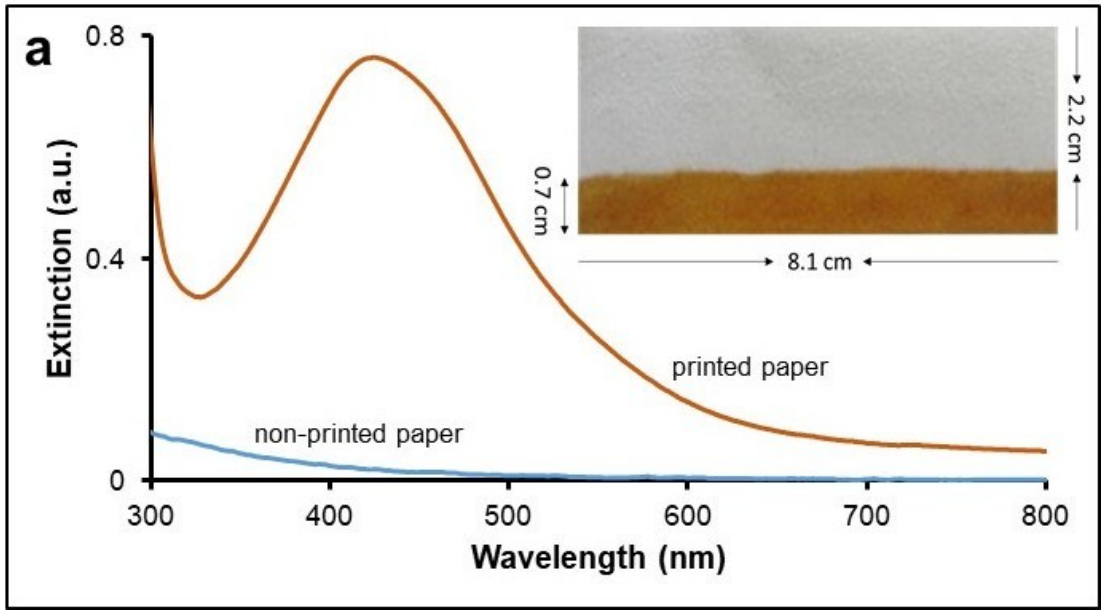

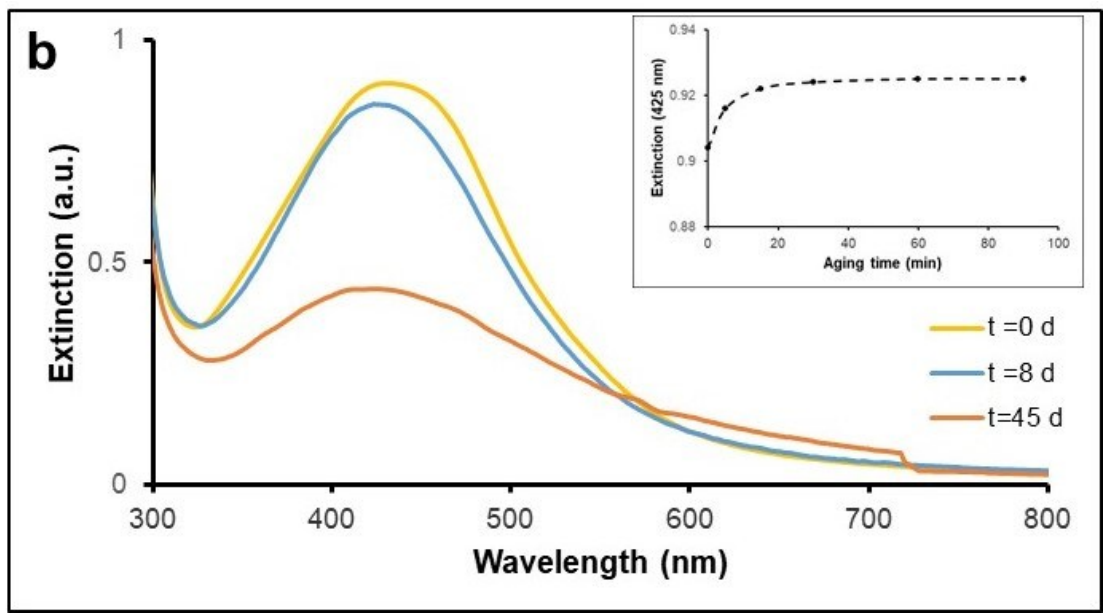

**Fig. 3**

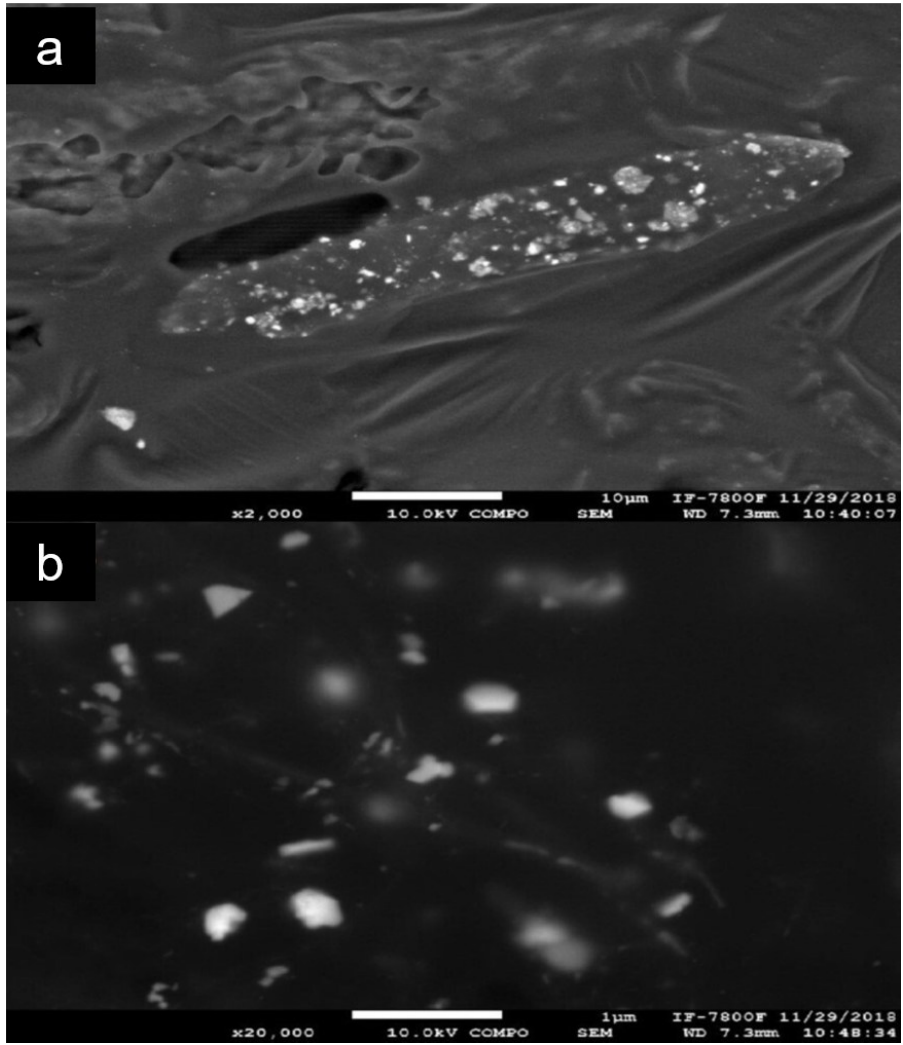

**Fig. 4**

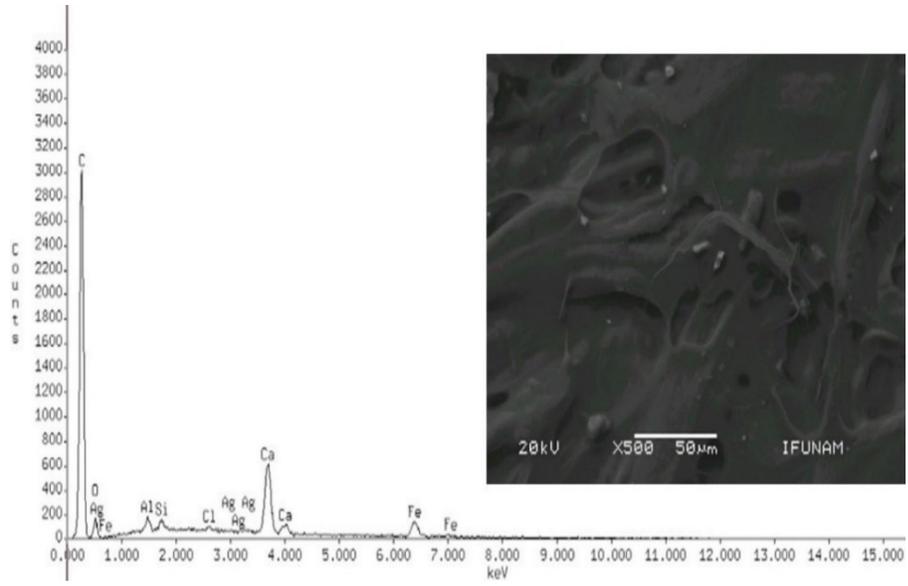

**Fig. 5**

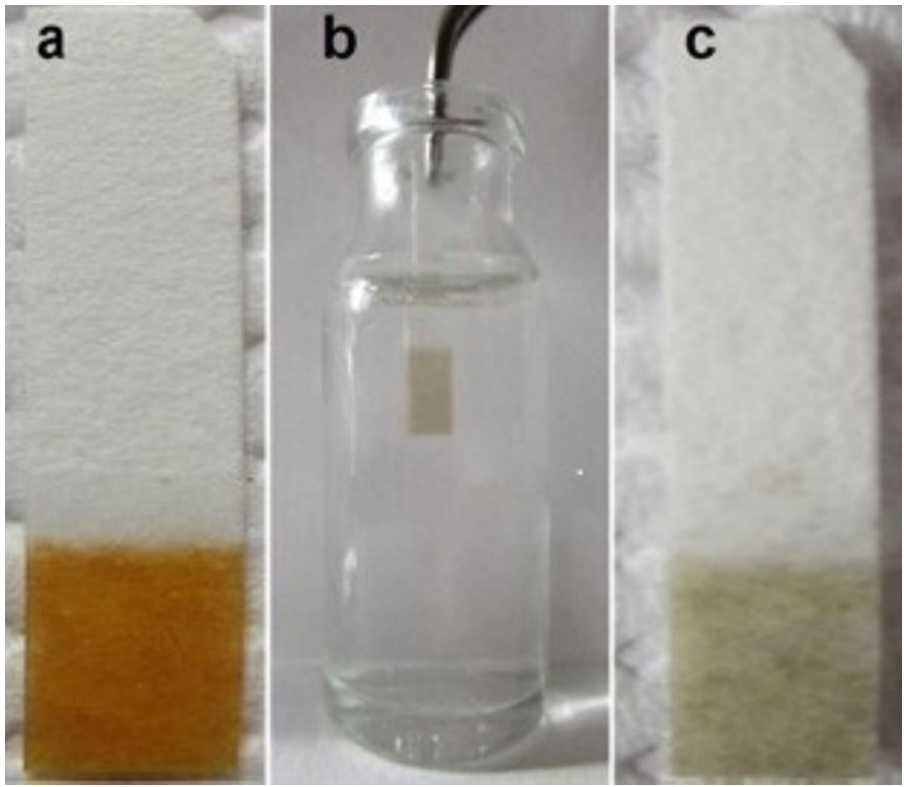

**Fig. 6**

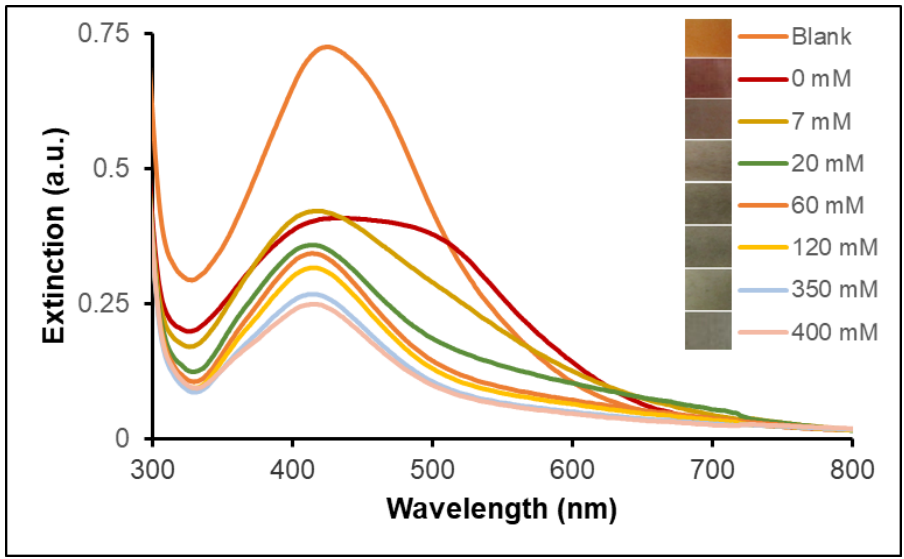

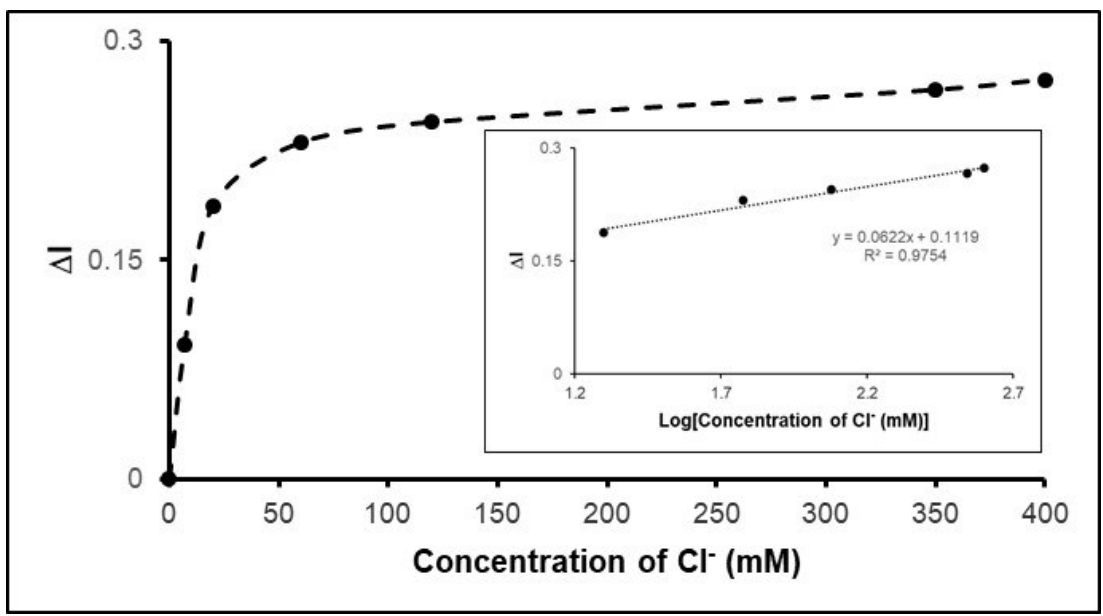

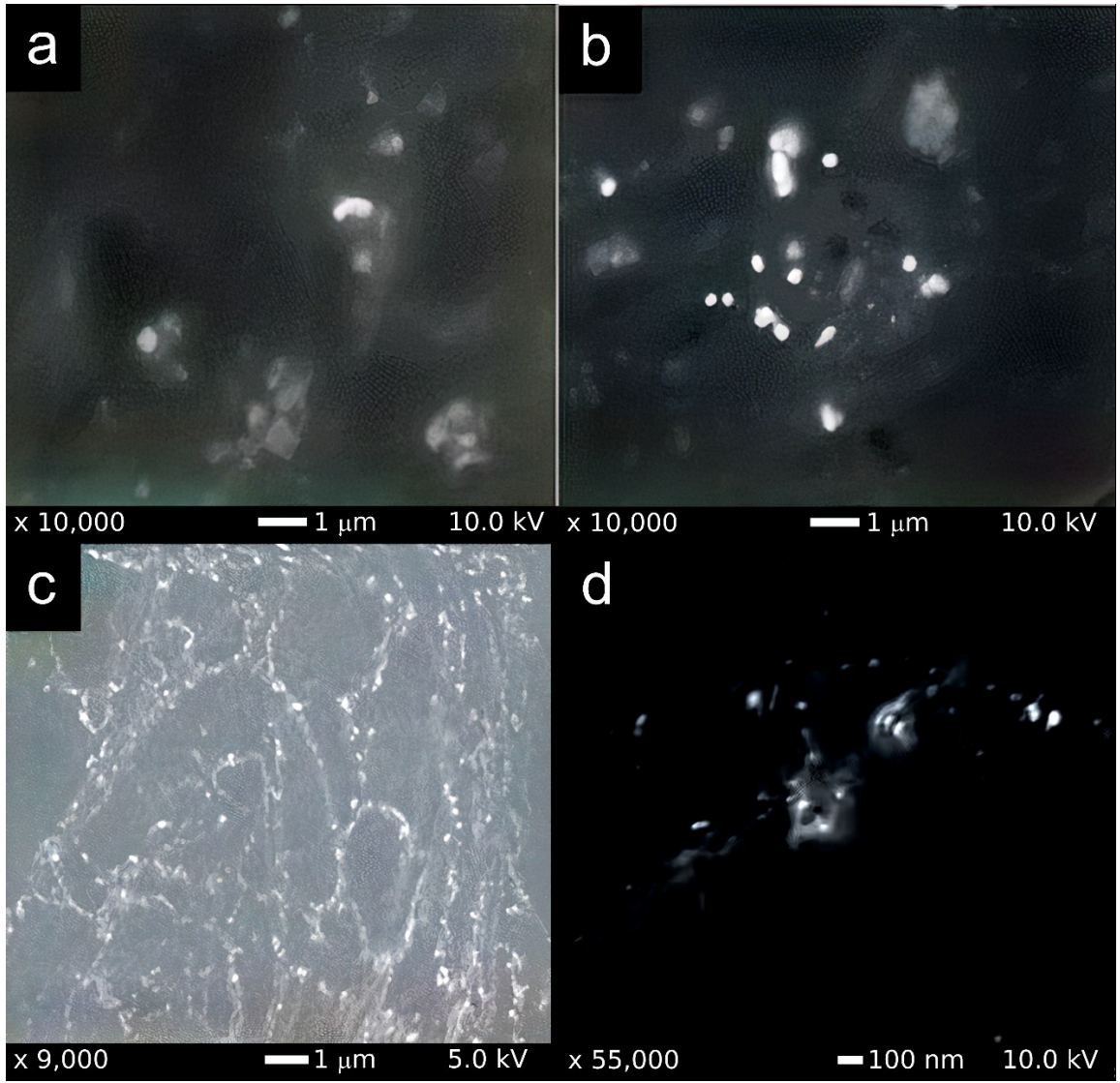

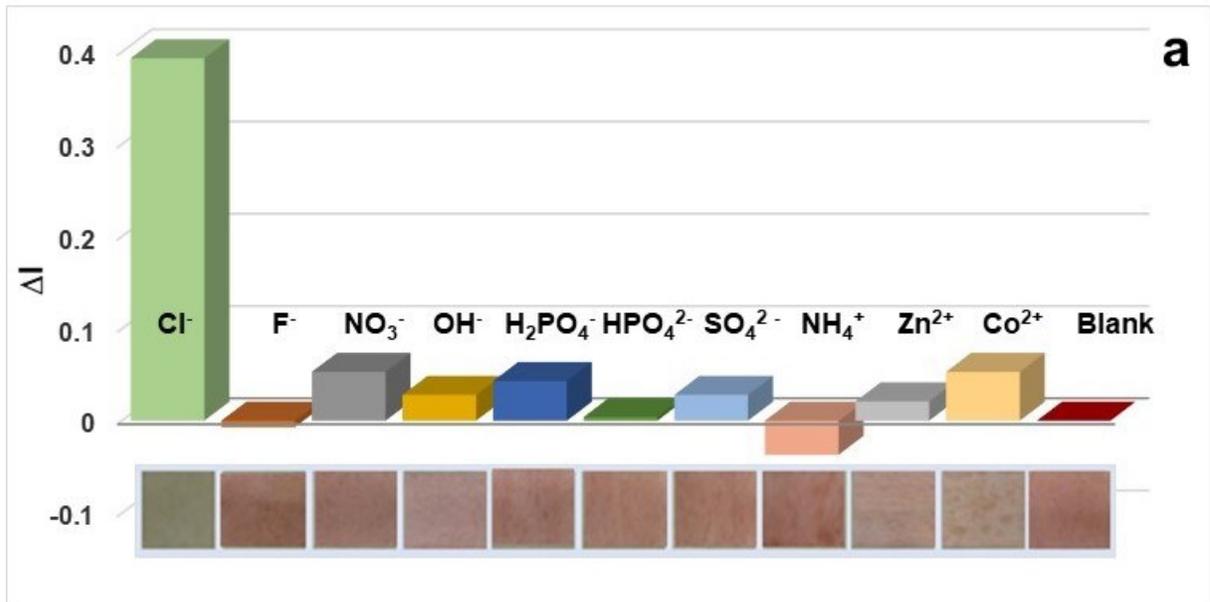

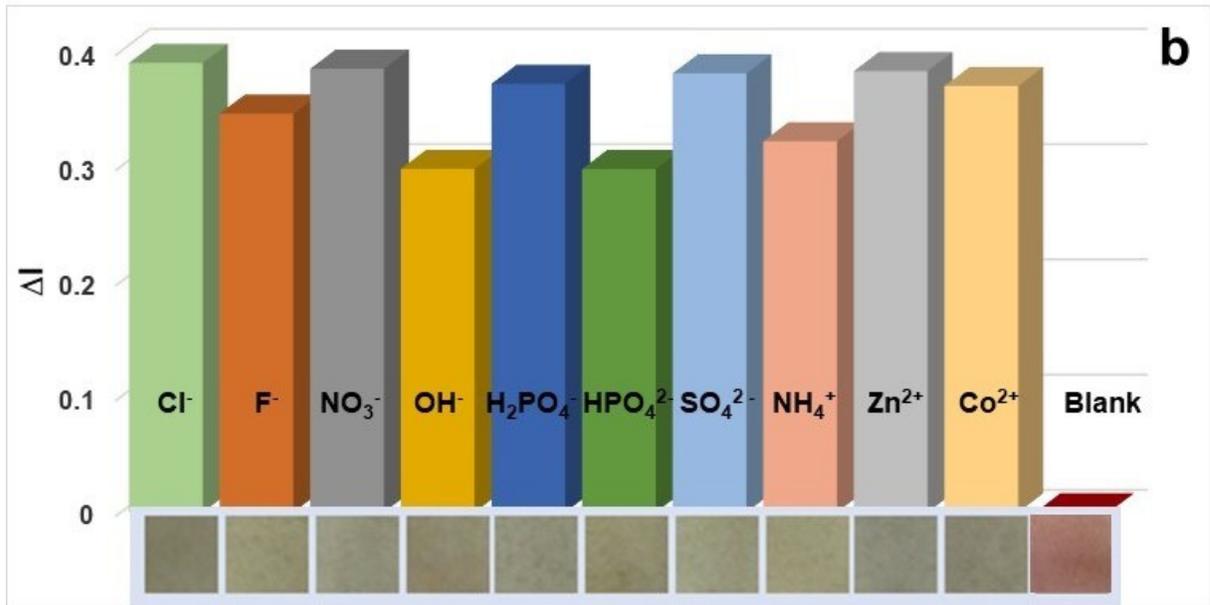

**Fig. 9c**

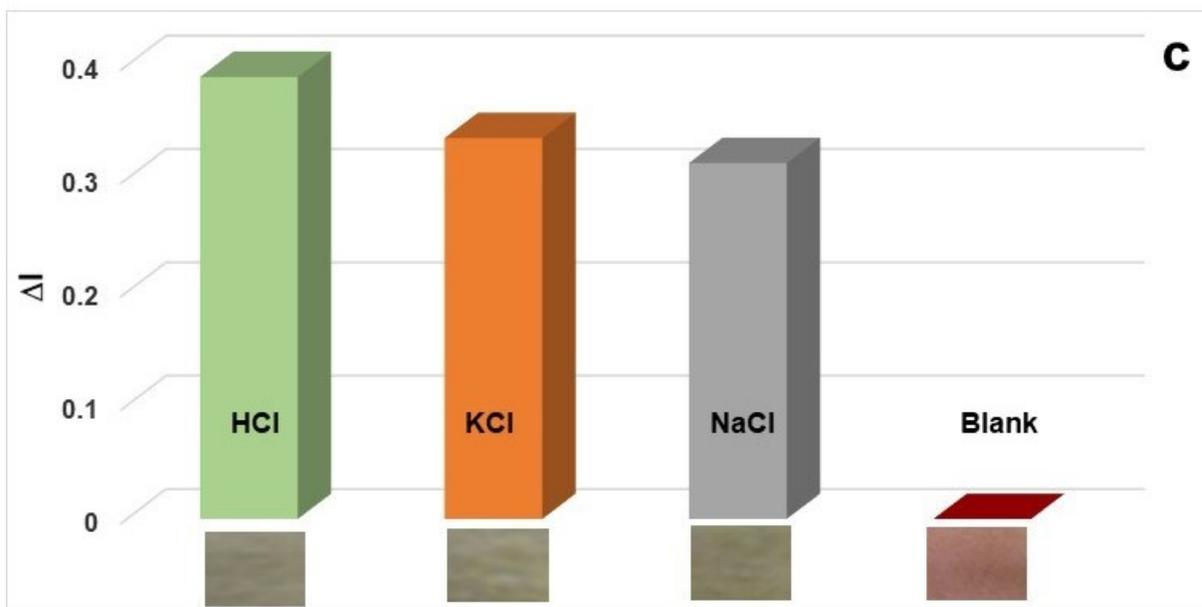

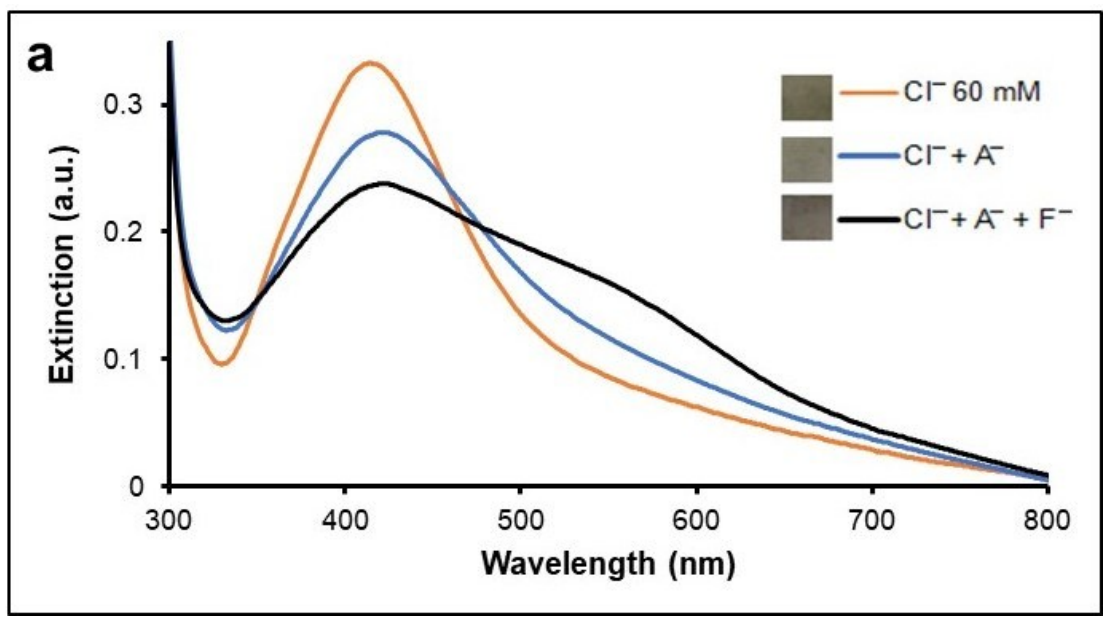

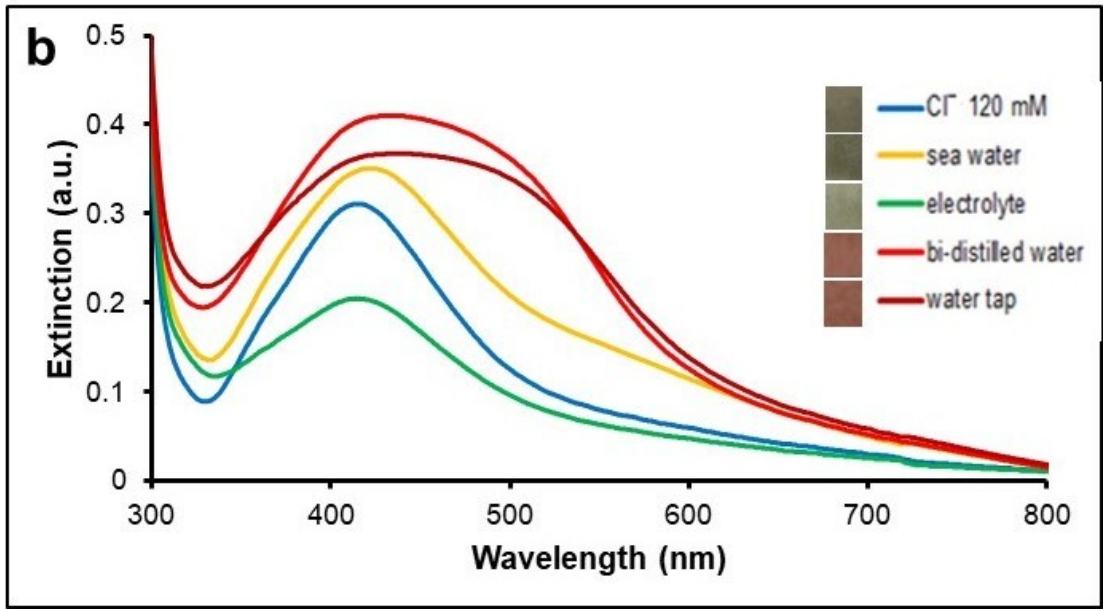

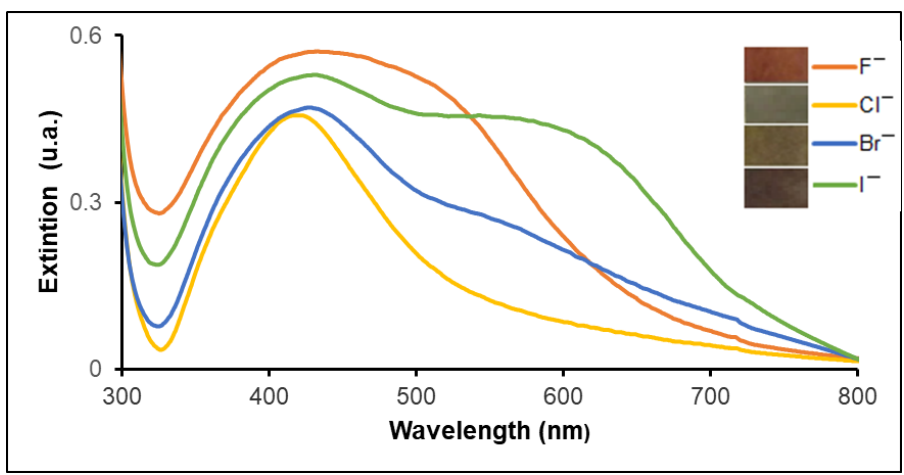

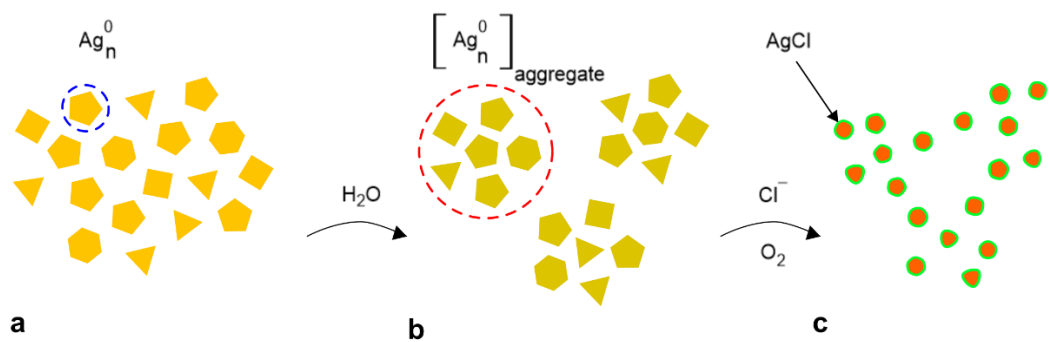

$$(4) \quad Ag^0 + Cl^- + O_2 + 2\,H_2O \longleftrightarrow AgCl + 4\,OH^-$$

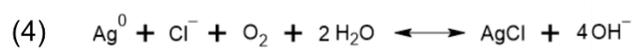

$$(5) \quad Ag_n^0(paper) + Cl^- + O_2 + 2\,H_2O \longrightarrow \left[ Ag_n^0(paper) \,@\, AgCl \right]_{aggregates} + 4\,OH^-$$

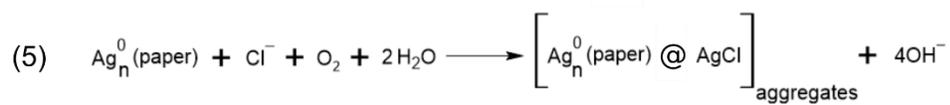